\documentstyle[preprint,aps,10pt]{revtex}
\begin{document}
\draft
\hfill\vbox{\baselineskip14pt
            \hbox{February 2001}}
\baselineskip20pt
\vskip 0.2cm 
\begin{center}
{\Large\bf EXAFS study of Tl$_{0.75}$Cu$_{.25}$Ba$_2$Ca$_3$Cu$_4$ O$_y$ 
 \\
 and Cu$_{0.68}$C$_{.32}$Ba$_2$Ca$_3$Cu$_4$ O$_y$ superconductors [bulk] 
 at 300 K}
\end{center} 
\vskip 0.2cm 
\begin{center}
\large Sher~Alam$^{1}$,~H.~Oyanagi$^{1}$, S.~B.~Mohamed$^{1}$,
~H.~Ihara$^{3}$, A.~Iyo$^{3}$, H.~Kito$^{3}$,~P.~Badica$^{3}$,
~M.~O.~Rahman$^{2}$ and T.~Yanagisawa$^{3}$

\end{center}
\begin{center}
$^{1}${\it Photonics, AIST, Tsukuba, Ibaraki 305, Japan}\\
$^{2}${\it  GUAS \& Photon Factory, KEK, Tsukuba, Ibaraki 305, Japan}\\
$^{3}${\it Nanoelectronics, AIST, Tsukuba, Ibaraki 305, Japan}
\end{center}
\vskip 0.2cm 
\begin{center} 
\large Abstract
\end{center}
\begin{center}
\begin{minipage}{14cm}
\baselineskip=18pt
\noindent

The results of EXAFS measurements at 300 K for the superconducting 
compounds Tl$_{0.75}$Cu$_{0.25}$Ba$_{2}$Ca$_{3}$Cu$_4$O$_{y}$ [Tl-1234], 
TlBa$_{2}$Ca$_{3}$Cu$_{4}$O$_{y}$ [Tl-1212], and 
CuBa$_{2}$Ca$_{3}$Cu$_{4}$O$_{y}$ [Cu-1234]. are reported. We have 
measured the EXAFS spectrum for 
Tl$_{0.75}$Cu$_{0.25}$Ba$_{2}$Ca$_{3}$Cu$_4$O$_{y}$
in the range 10K-300K, however here we limit our discussion to
the spectrum at 300 K. This material is prepared under high 
pressure [3.5 GPa] from precursors with small carbon concentrations 
and exhibits a T$_c$ of $~127$ K. We have also performed ``aging'' study 
by looking at XRD for this material after approximately one year.
The XRD results at 300 K are ``unchanged''. It is of
interest to compare the EXAFS spectrum of this compound with the  
corresponding compound Cu-1234. Remarks on the choice of 
appropriate EXAFS standard for this and related compounds
are also given. Based on our data analysis we quantify disorder
in these systems. By using the Cu-O in-plane distance we give
values for the microstrain parameter, which can be related to
the charge ordering transition.


\end{minipage}
\end{center}
\vfill
\baselineskip=20pt
\normalsize
\newpage
\setcounter{page}{2}
\section{Introduction}
	This one of a series of papers where we report on the EXAFS 
XANES spectra and XRD of the TlBa$_{2}$Ca$_{n-1}$Cu$_{n}$O$_{y}$ 
[n=2,3,4,5], [Tl-12n(n+1)] and CuBa$_{2}$Ca$_{n-1}$Cu$_{n}$O$_{y}$ 
[n=2,3,4,5], [Cu-12n(n+1)]\cite{iyo00}. Here we mainly concentrate on 
the EXAFS spectra of TlBa$_{2}$Ca$_{3}$Cu$_{4}$O$_{y}$ [Tl-1234],
CuBa$_{2}$Ca$_{3}$Cu$_{4}$O$_{y}$ [Cu-1234], 
TlBa$_{2}$Ca$_{2}$Cu$_{3}$O$_{y}$ [Tl-1223] and 
TlBa$_{2}$Ca$_{1}$Cu$_{2}$O$_{y}$ [Tl-1212] at room temperature. 
One of our purpose is to establish good theoretical and 
experimental ``standards''
for these systems. We assume for the purposes of this note 
the experimental standards to be Tl-1212 and Tl-1223. For
the Cu-12n(n+1) compounds it would be reasonable to choose
Cu-1212 and Cu-1223 with appropriate nominal compositions
as standards. However we have not yet measured the spectra for 
Cu-1212 and Cu-1223. Thus we tentatively utilize Tl-1212 and
Tl1223. The theoretical standards are calculated with
FEFF 8.10. However all the calculations are preliminary. 

	From the studies of these families it is generally 
concluded that both are isostructural to each other except for the 
position of oxygen in the charge reservoir layer. It is also 
assumed that Tl affects the synthesis parameters and by changing
its valence state varies the number of carriers and oxygen content
in the conventional Tl-based superconductors \cite{mor97}.

	For the family TlBa$_{2}$Ca$_{n-1}$Cu$_{n}$O$_{3+2n}$
the n=2 [i.e. Tl-1212] and  n=3 [i.e. Tl-1223] we have tentatively
used the refined single crystal x-ray diffraction data reported 
in \cite{her94} [Tables 6 and 8 page 30 of \cite{her94}] for our
theoretical calculations performed by FEFF 8.10. The listed 
crystallographic data [at room temperature] is for Tl-1212 
[nominal composition [TlBa$_{2}$(Ca$_{0.83}$Tl$_{0.17}$)Cu$_{2}$O$_{6.75}$]
the $T_c= 103$ K, a= 3.8566(4) \AA,  and c= 12.754(2) \AA.
The listed crystallographic data [at room temperature] is for Tl-1223 
[nominal composition [TlBa$_{2}$(Ca$_{0.95}$Tl$_{0.05}$)$_{2}$
Cu$_{2}$O$_{6.75}$] the $T_c= 103$ K, a= 3.853(1) \AA,  and 
c= 15.913(4) \AA. For this family [TlO monolayer] compounds the 
symmetry groups is primitive tetragonal P4/mmm. We note that our 
sample Tl-1212 has a nominal composition 
Tl$_{0.75}$ Cu$_{0.25}$ Ba$_{2}$ Ca Cu$_{2}$ O$_{7}$, and the 
$T_c= 100$ K. The sample Tl-1234 has a nominal composition 
Tl$_{0.75}$ Cu$_{0.25}$ Ba$_{2}$ Ca$_{3}$ Cu$_{4}$ O$_{11}$, 
with $T_c= 127$ K and lattice parameters  a= 3.848 \AA,  
and c= 19.08 \AA. The sample Cu-1234 has a nominal composition 
Cu$_{0.68}$ C$_{0.32}$ Ba$_{2}$ Ca$_{3}$ Cu$_{4}$ O$_{11}$, 
with $T_c= 117$ K, the lattice parameters are  a= 3.859 \AA, and 
c= 17.95 \AA.

	Previous EXAFS study \cite{all91} claims local
structural changes [for Tl-1234] occuring around T$_{c}$ 
and pair distribution function (PDF) analysis of pulsed neutron
scattering \cite{tob90} also suggests that electron-lattice interaction
may be involved in the mechanism of superconductivity.
In contrast no discontinuity or significant structural
change is observed through T$_{c}$ in refined parameters
for each of the compounds Tl-2201 and Tl-2223. Thus this
was one of our motivation to take a look at the temperature
dependent EXAFS, XANES and XRD study of Tl-1234. However
we must keep in mind that EXAFS is sensitive to local
dynamical changes on the order of 10$^{-15}$ s and hence
could detect a phonon anomaly around T$_{c}$ whereas XRD
cannot probe such a change.	

	The purpose of this note is to report the results
of measurements of Tl-1234 system, and also of Tl-1212,
Tl-1223 and Cu-1234. We have chosen this set since we want
to create an appropriate experimental and theoretical standard
for the family TlBa$_{2}$Ca$_{n-1}$Cu$_{n}$O$_{y}$ [n=2,3,4]
and related materials. To this end we have worked within
this family, and Cu-1234 since it is isostructural to 
Tl-1234\footnote{Cu-1234 has several advantages \cite{aki97}}. 
In particular we try to self-consistently study the Cu-O bond, both
in-plane and apical. Yet another motivation is to quantify
disorder within this system. Once the Cu-O bond is determined
one can make a prediction for charge ordering transitions.

	The set-up of this paper is as follows, in the 
next section we outline some experimental details,
section three deals with results and discussion and finally
we state the conclusions. 
\section{Experimental}
\subsection{Sample}
	The samples TlBa$_{2}$Ca$_{n-1}$Cu$_{n}$O$_{y}$ [n=3,4]
with $T_{c}=133.5$ K for n=3 and $T_{c}=127$ K for n=4 [after
post-annealing under nitrogen flow] have been syntheisized under 
a pressure of 3.5 GPa using precursors with small residual carbon 
concentrations by A.~Iyo et al. \cite{iyo00}. It is suggested that
the supression of the substitution of Tl for Ba and Ca sites is
mainly responsible for the $T_{c}$ enhancement \cite{iyo00}. Details
of sample preparation can be found in \cite{iyo00}. The sample was
heated at 850 $^{o}$C $\sim$ 950 $^{o}$C for 2 hours under a pressure
of 3.5 GPa using a cubic-anvil-type apparatus [Riken CAP-07].
The as synthesized Tl-1223 and Tl-1234 exhibited over-doping state,
and T$_c$ of 100 K and 120 K respectively. Thus post-annealing under 
nitrogen gas was employed to enhance the doping state by reducing the
oxygen content. Annealing time depends on annealing temperature
and the following sequence was employed 650 $^{o}$C for 6-12 hours,
675 $^{o}$C for 1-2 hours or 700 $^{o}$C for 0.5-1 hours for
polycrystalline samples of approximate mass 50 mg. Thus a 
post-annealing at temperatures about 50 $^{o}$C above 650 $^{o}$C
are optimal for $T_{c}$ enhancements. We have performed 
x-ray diffraction [XRD] measurement on Tl-1234 after approximately
one year after it was prepared and compared to the XRD results 
obtained just after preparation, and found no change.  
\subsection{EXAFS Measurements}
	The EXAFS and XANES measurements were conducted at
beamline B13-B1 at the Photon Factory (PF), Tsukuba, in the
Fluorescence mode. The electron beam energy was 2.5 GeV and 
the maximum stored current is 400 mA. EXAFS data was collected
using fixed-exit double crystal Si(111) monochromator. The
first crystal is a water-cooled flat Si(111) monochromator
and the second crystal is sagitally bent to focus the
horizontal beam over $\equiv$ 2 mrad, Fig.~\ref{fig1}. 
A 19-element solid state detector was used to collect the 
flourescence signal. The large number of detectors allows us 
to cover a sizeable amount of the solid angle of the x-ray 
flourescence emission in addition to giving a high signal 
to noise ratio. 
\section{Results and Discussion}
As is known the EXAFS amplitude can be written as
\begin{eqnarray}
\chi(k)=\frac{m \pi}{h^2}\sum_{i}3 N_{i}\cos^{2}(\theta_i)
[S_{0}^{2}/(k R_{i}^{2})]f_{i}(k,R_i)e^{-2R_i/\lambda)}
e^{-2k^{2}\sigma_i^{2}}\sin[2 k R_i + \delta_i].
\label{r1}
\end{eqnarray}
The value $N_{i}$ labels the equivalent number of neigboring
atoms to the absorbing atoms, at a distance $R_i$ and angle
$\theta_i$ with respect to the polarized electric field vector
of synchrotron radiation. The backscattering amplitude is
$f_{i}(k,R_i)$, the photoelectron mean free path is $\lambda$,
the correlated Debye-Wallar factor of the absorbing-backscattering
atoms is called $\sigma_i^{2}$, $S_{0}^{2}$ [NF] represents the
amplitude correction factor due to the electron correlations
and $\delta_i$ is the value of phase shift.
The oscillatory part $\chi(k)$ in Eq.~\ref{r1} is defined relative
to smooth atomic background absorption $\mu_{0}(E)$,
\begin{eqnarray}
\chi = (\mu -\mu_0)/\mu_0.
\label{r2}
\end{eqnarray}
The Fourier transform [FT] of the k-space [oscillatory] data
produces peaks in real space [r-space]. These peaks 
correspond to neighbours at various radial distances
from the absorbing atom, or the central atom.

	The experimental raw data is converted to $\chi(k)$
using Ada1\footnote{A XAFS program developed by H.~Oyanagi
to normalize, subtract EXAFS and XANES data and to produce
$\chi(k)$ from the EXAFS data}. We note that we the 
$E_0$ value we have used in Ada1 is $8.9803$ keV.
The $\chi(k)$ obtained by using Ada1 is then fed 
into Ada2.5\footnote{A XAFS program developed by H.~Oyanagi}.
The numerical values for $R_i$, $N_i$ and $\sigma_{i}$
are obtained by performing iterated least-square fits
of the real and imaginary parts of the FT of $\chi(k)$
to a FEFF6.01a generated theoretical 
data\footnote{Unfortunately Ada2.5 has not been updated
to accept current FEFF8.10 input.}. As it is known, a
common problem in XAFS data analysis is too many fitting
parameters if the fit includes further shells in a
complicated system. For example, if we want
to fit in the k-space range $3.0 \sim 13.0$~\AA$^{-1}$~
and take r-space as $1.0 \sim 4.0$~\AA~we obtain as an estimate 
$2(\Delta k)(\Delta r)/\pi+2 \approx 21$ parameters
to be fitted. In a typical system such as YBCO one needs
24 [assuming 8 pairs] or more fitting parameters, even if
fix the edge energy. Thus a constrained fit, in general, is 
required in XAFS data analysis.

	Figs.~\ref{fig2}-\ref{fig5} show the raw data
obtained as  result of our measurements for Tl-1234,
Tl-1212, Tl-1223 and Cu-1234 respectively. 
Figs.~\ref{fig6}-\ref{fig10} are the $\chi(k)$ results 
of theoretical calculations using FEFF8.10, for
Tl-1234, Tl-1212, Tl-1223, Cu-1234 and Cu-1234 [with
and without] calcium respectively. The structure near
$k \approx 0-2.5$~\AA$^{-1}$ in Fig.~\ref{fig6} for
Tl-1234 is due mainly due to thallium, we verified
this by excluding thallium and found that the peak
near $k \approx 0.5-1.0$~\AA$^{-1}$ disappeared.
From our raw data we can naively say that data
quality of Tl-1212 relative to Tl-1223, Tl-1234
and Cu-1234 is slightly inferior and must be
improved.    

	The following results were obtained using Ada2.5.
We want to get an estimate for `static' disorder for
TlBa$_{2}$Ca$_{n-1}$Cu$_{n}$O$_{y}$ [n=2,3,4] family
and Cu-1234 associated with Cu-O in-plane bond.
We obtain the following Cu-O in-plane results for
the various systems. For Tl-1212
Cu-O(1)\footnote{To be consistent we use the notation of
the cited paper} distance from XRD is 1.929 \AA \cite{her94}. 
We obtain $N=3.78$, $r=1.929$~\AA,
$\sigma=0.08172$, $\delta E_{0}=5$ eV and NF=0.82435.
Values for Tl-1223 are: $N=3.77$, $r=1.927-1.9265$~\AA,
$\sigma=0.08572$, $\delta E_{0}=5$ eV and NF=0.82435.
In the case of Tl-1234 the values are:  $N=3.76$, $r=1.908$~\AA,
$\sigma=0.095$, $\delta E_{0}=5$ eV and NF=0.82435. The
Cu-1234  the values are:  $N=3.76$, $r=1.927$~\AA,
$\sigma=0.0914$, $\delta E_{0}=5$ eV and NF=0.82435. The
$r$ value which gives the best fit in case of Cu-1234
is the same value quoted in \cite{aki97} for nominal
composition Cu$_{0.68}$ C$_{0.32}$ Ba$_{2}$ Ca$_{3}$ Cu$_{4}$ O$_{11}$,  
which is the same composition of the sample we used
in our measurements. This provides a crude check on our
analysis. Since all the materials were
measured at the approximately same temperature [i.e. 300 K],
we can see that the change in $\sigma$ can be related to
static disorder. Thus taking the Tl-1212 as a standard,
we can say that there is approximately a 4.89\% change
in $\sigma$ in Tl-1223, a 16.25\% change in $\sigma$ in 
Tl-1234 and roughly a 11.85\% $\sigma$ increase in Cu-1234. 
This implies [roughly] that the Cu-O in plane bond is associated
with greater static disorder in Tl-1234 compared
to Cu-1234. 

	Next we consider the Cu-O apical bond. We assume that
DWF is approximately unchanged from the Cu-O in-plane result.
The following results are obtained: For Tl-1212
$N=1.0$, $r=2.89$~\AA, $\sigma=0.08172$, $\delta E_{0}=5$ eV and 
NF=0.82435. In case of Tl-1223 $N=1.0$, $r=2.85$~\AA, $\sigma=0.085$, 
$\delta E_{0}=5$ eV and NF=0.82435. For Tl-1234 the values 
are: $N=0.26$, $r=2.76$~\AA, $\sigma=0.095$, $\delta E_{0}=5$ eV and 
NF=0.82435. In case of Cu-1234 $N=0.86$, $r=2.45$~\AA, $\sigma=0.0914$, 
$\delta E_{0}=5$ eV and NF=0.82435.

	One of our motivation for studying the EXAFS, XANES
and XRD of cuprates is to understand the electron-lattice
interactions and their possible role [if any] in the 
mechanism of superconductivity in these materials. Yet
another factor is to understand stripes and more generally 
charge ordering. A possible parameter which may quantify charge 
ordering is microstrain. We define [a la A. Bianconi\cite{bia00}] 
the microstrain parameter for the Cu-O bond $\epsilon$, where
\begin{equation}
\epsilon=2(1-R_{_{Cu-O}}/d_0)
\label{e1}
\end{equation}
here $d_0 \equiv 1.985\pm 0.05$ \AA, which is taken an indication
if a system will develop a charge order.
EXAFS experiments seem to indicate that certain materials may
not show stripe phase if the microstrain $\epsilon$ is less than
a certain critical value $\equiv 0.045$. Putting in the values
given in \cite{all91} $R_{_{Cu-O}}= 1.901,~1.904,~1.907,~1.907,
~1.912,~1.904$ \AA~at 10 K, 100 K, 117 K, 127 K, 135 K, 156 K
respectively in Eq.~\ref{e1} we obtain the values 0.08463, 0.08161,
0.078589, 0.078589, 0.073552, and 0.08161 for $\epsilon$.
Our value of $R_{_{Cu-O}}$ is $1.908$\AA, thus 
$\epsilon=0.077582$. Since these numbers are bigger than 
the critical microstrain value, we naively expect a charge 
ordering transition. For Tl-1212 we obtain 
$\epsilon=0.056423$, and in case of Cu-1234 we get
$\epsilon=0.075567$.
\section{Conclusions}
We have given the results of our measurement of the EXAFS
spectra of the three compounds Tl-1234, Tl-1212 and Cu-1234.
The oscillatory part $\chi(k)$ of the measured data is
determined using the program Ada1. The theoretical $\chi(k)$ 
have been calculated for Tl-1234, Tl-1212, Tl-1223 and Cu-1234
using FEFF8.10. Using Ada2.5 we have given preliminary results
for Cu-O in-plane and apical bonds for the thallium compounds
Tl-1212, Tl-1223 and Tl-1234 and also for Cu-1234. From our
estimation of $\sigma$ we have conclude that Tl-1234 is
more disordered than Tl-1212 and Cu-1234. Using the simple
criterion for predicting charge ordering transition we can
see that Tl-1212, Tl-1223, Cu-1234, and Tl-1234 satisfy
this condition. To obtain a self-consistent standard for 
the thallium and copper families we have to get many more 
EXAFS  measurements and combine it with both XRD and neutron 
measurements. This can quantify charge ordering in these materials
and also can be used in combination with electron
diffraction to map out charge density distribution
in detail associated with Cu-O bonds.
\section*{Acknowledgments}
The Sher Alam's work is supported by the Japan Society for
for Technology [JST]. S.~B.~Mohamed's work is supported by 
the New Energy and Industrial Technology Development 
Organization (NEDO) as Collaborative Research and Development of
Fundamental Technologies for Superconductivity Applications.


\begin{figure}
\caption{Schematic diagram of the experimental set-up}\label{fig1}
\end{figure}
\begin{figure}
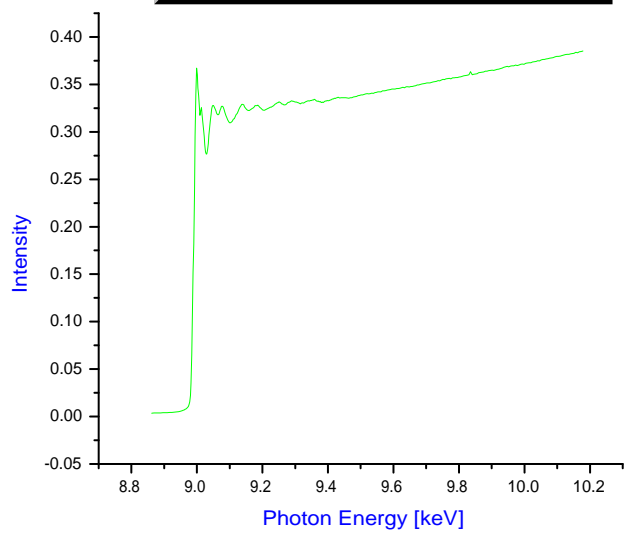

\caption{Raw EXAFS data for Tl-1234 at 300 K}\label{fig2}
\end{figure}
\begin{figure}
\caption{Raw EXAFS data for Tl-1212 at 304-305 K}\label{fig3}
\end{figure}
\begin{figure}
\caption{EXAFS data for Tl-1223 at 300 K annealed in nitrogen
for 1 hour at tempearture of 675 $^{o}$C}\label{fig4}
\end{figure}
\begin{figure}
\caption{Raw EXAFS data for Cu-1234 at 304-305 K}\label{fig5}
\end{figure}
\begin{figure}
\caption{$\chi(k)$ calculated with FEFF 8.10 for Tl-1234 system}\label{fig6}
\end{figure}
\begin{figure}
\caption{$\chi(k)$ calculated via FEFF 8.10 for Tl-1212}\label{fig7}
\end{figure}
\begin{figure}
\caption{$\chi(k)$ calculated with FEFF 8.10 for Tl-1223}\label{fig8}
\end{figure}
\begin{figure}
\caption{$\chi(k)$ calculated using FEFF 8.10 with FEFF8.10 with
calcium assuming $R_{{\rm max}}=6.5$~\AA~for 
 Cu$_{0.68}$ C$_{0.32}$ Ba$_{2}$ Ca$_{3}$ Cu$_{4}$ O$_{11}$}\label{fig9}
\end{figure}
\begin{figure}
\caption{$\chi(k)$ calculated using FEFF 8.10 with FEFF8.10 
showing the result with [red] and without [green] calcium for 
 Cu$_{0.68}$ C$_{0.32}$ Ba$_{2}$ Ca$_{3}$ Cu$_{4}$ O$_{11}$}\label{fig10}
\end{figure}


\begin{references}
\bibitem{iyo00}A.~Iyo et al., 13th International Symposium on 
Superconductivity [ISS 2000], and references therein.
\bibitem{mor97}B.~Morosin et al., Physica~C~{\bf 288},
~(1997)~255-267.
\bibitem{her94}Allen~M.~Hermann and J.~V.~Yakhmi eds.,
Thallium-Based High-Temperature Superconductors, Marcel
Dekker, Inc., 1994.
\bibitem{all91}P.~G.~Allen~et al., PRB~{\bf 44},(1991),9480-9485.
\bibitem{tob90}B.~H.~Toby~et al., PRL~{\bf 64},(1990),~2414-2417.
\bibitem{aki97}J.~Akimoto et al., Physica~C~{\bf 279},
~(1997)~181-196.
\bibitem{bia00}A.~Biaconni~et al.,~J.~Phys.:~Condens.~Matter.~{\bf 12}
,(2000),~10665.
\end{references}
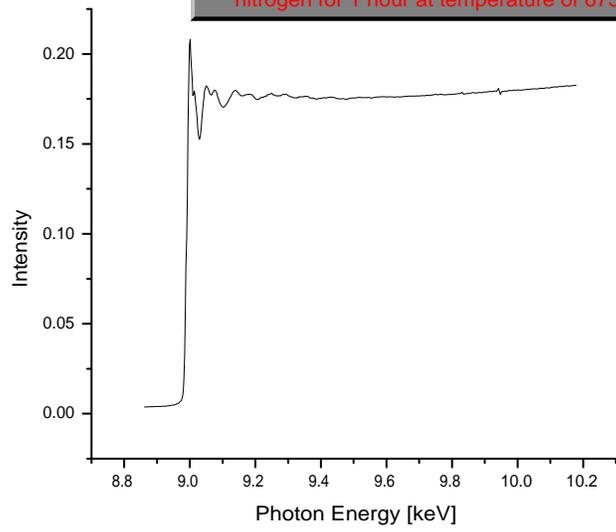
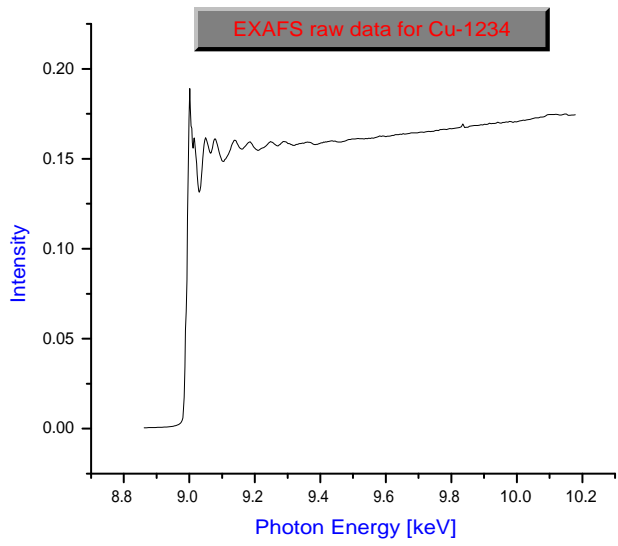
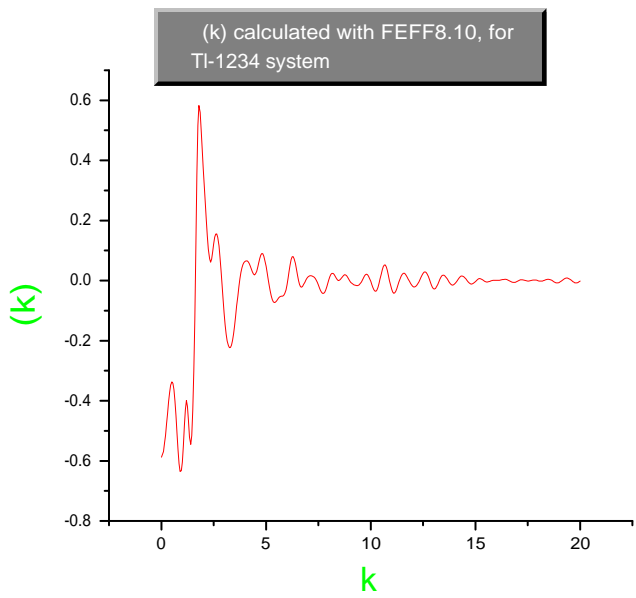
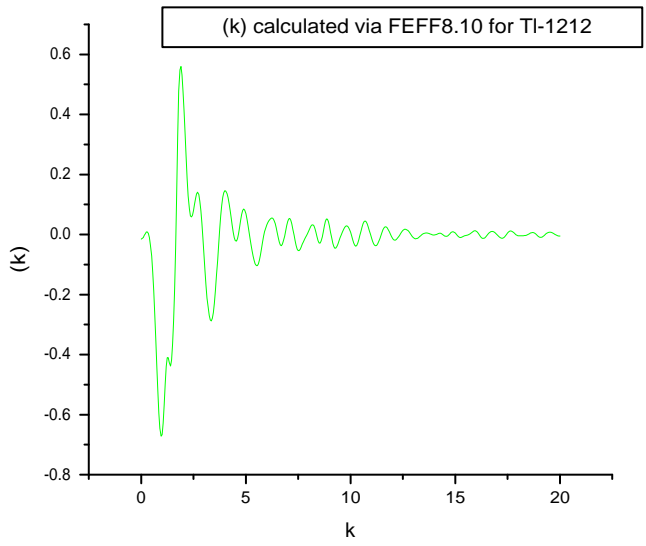
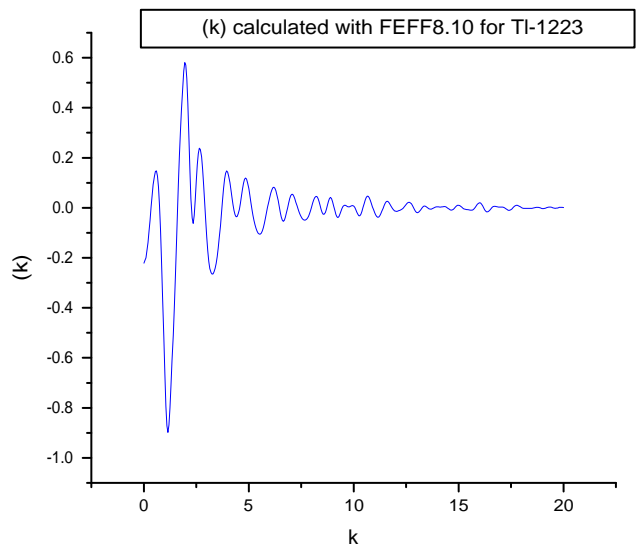
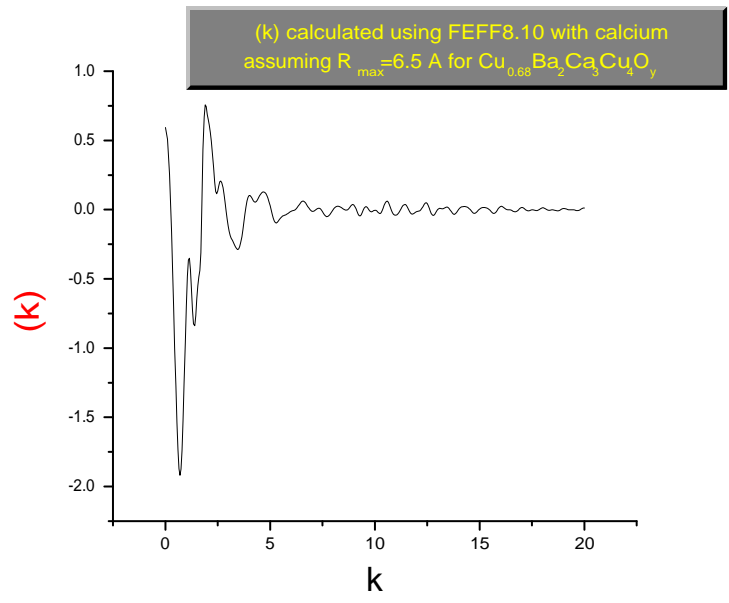
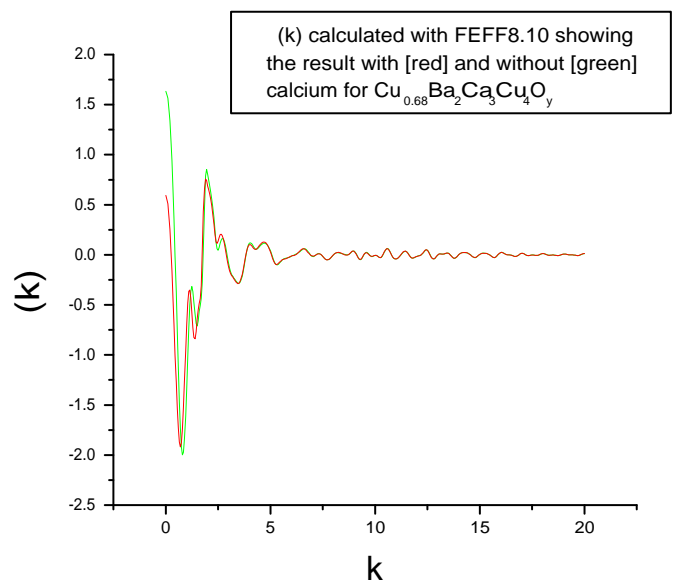
\end{document}